\documentclass[floats,floatfix,showpacs,preprint,prd,superscriptaddress,onecolumn,nofootinbib]{revtex4}
\usepackage{graphicx, epsfig, amssymb} 
\usepackage{amsmath, amsfonts}
\usepackage{bm} 
\usepackage{color}
\usepackage[usenames]{xcolor}

\newcommand{\be}{\begin{equation}}
\newcommand{\ee}{\end{equation}}                  
\newcommand{\bea}{\begin{eqnarray}}
\newcommand{\eea}{\end{eqnarray}}

\begin{document}

 
\title{Pleba\'nski-Demia\'nski solutions in Quadratic gravity with conformally coupled scalar fields}


\author{Adolfo Cisterna}
\email{adolfo.cisterna@ucentral.cl}
\affiliation{Dipartimento di Fisica, Universita di Trento, Via Sommarive 14, 38123 Povo (TN), Italy.}

\affiliation{TIFPA - INFN, Via Sommarive 14, 38123 Povo (TN), Italy.}

\author{Anibal Neira-Gallegos}
\email{aneira2017@udec.cl}
\affiliation{Departamento de F\'isica, Universidad de Concepci\'on, Casilla, 160-C, Concepci\'on, Chile.}

\author{Julio Oliva}
\email{juoliva@udec.cl}
\affiliation{Departamento de F\'isica, Universidad de Concepci\'on, Casilla, 160-C, Concepci\'on, Chile.}

\author{Scarlett C. Rebolledo-Caceres}
\email{srebolledo2017@udec.cl}
\affiliation{Departamento de F\'isica, Universidad de Concepci\'on, Casilla, 160-C, Concepci\'on, Chile.}

\begin{abstract}
We show that the Pleba\'nski-Demia\'nski spacetime persists as a solution of General Relativity when the theory is supplemented with both, a conformally coupled scalar theory and with quadratic curvature corrections. The quadratic terms are of two types and are given by quadratic combinations of the Riemann tensor as well as a higher curvature interaction constructed with a scalar field which is conformally coupled to quadratic terms in the curvature. The later is built in terms of a four-rank tensor $S_{\mu\nu}^{\ \ \lambda\rho}$ that depends on the Riemann tensor and the scalar field, and that transforms covariantly under local Weyl rescallings.
Due to the generality of the Pleba\'nski-Demia\'nski family, several new hairy black hole solutions are obtained in this higher curvature model. We pay particular attention to the C-metric spacetime and the stationary Taub-NUT metric, which in the hyperbolic case can be analytically extended leading to healthy, asymptotically AdS, wormhole configurations. Finally, we present a new general model for higher derivative, conformally coupled scalars, depending on an arbitrary function and that we have dubbed Conformal K-essence. We also construct spherically symmetric hairy black holes for these general models.

\noindent 

\end{abstract}

\maketitle


\section{Introduction}

In gravity, described by General Relativity (GR) or in any of its extensions, a fundamental angle to be investigated is their spectrum of solutions. Such set includes gravitational waves \cite{TheLIGOScientific:2017qsa} and black holes as well as other astrophysically relevant objects. Black holes are formed by gravitational collapse \cite{Joshi:2012mk}, a catastrophic event that may produce singularities, pathological points in the spacetime where geodesic completeness is broken and where the curvature may diverge \cite{Hawking:1973uf}. The classical validity of GR is guaranteed by the cosmic censorship hypothesis \cite{Penrose:1969pc} which asserts that under generic conditions and for realistic forms of matter the singularities are covered by a horizon, a surface from which no information can escape to infinity. Ultimately, a black hole is the region inside the event horizon, a region that is causally disconnected with future null infinity and whose boundary in the spacetime is given by the event horizon itself. 
It has been argued that, after the gravitational collapse takes place, the formed black hole can only be described by a well-defined set of parameters, its mass, electromagnetic charges and angular momentum, and that no possible further relevant characteristics of the original matter that produced the black hole, such as baryonic or leptonic numbers for example, survive the process. In some case this conjecture can be proven and leads to no-hair theorems \cite{Ruffini:1971bza}, where hair is used to name all possible characteristics of black holes that make them not bald, those quantities that do not hold a Gauss law, and as a consequence are not conserved at infinity. \\ 
In spite of these results, it is possible to construct hairy black holes in General Relativity with a sensible matter content. The first attempts in this direction were carried out with the backreaction of a conformally coupled scalar field \cite{BBMB,Bekenstein:1974sf}. A backreacting, minimally coupled, static, spherically symmetric massless scalar field inevitably leads to a singular spacetime, which can be integrated analytically \cite{Janis:1970kn,Janis:1968zz}. Such pathology remains even beyond spherical symmetry in the static case \cite{Bekenstein:1998aw}\footnote{Recently these no-hair results have been revisited in the context of stationary, axisymmetric spacetimes endowed with reasonable matter fields that do not share the same symmetries of the metric \cite{Canfora:2013osa,Herdeiro:2014goa}.}. Nevertheless, the introduction of a conformal coupling with the curvature allows to by-pass the no-hair results providing a healthy spacetime, but with a scalar field that diverges at the horizon. Notwithstanding this result, as shown in \cite{Bekenstein:1975ts} such singular behavior does not imply infinite tidal forces on a falling object. The presence of a cosmological constant and a quartic self-interaction pushes the scalar field singularity inside the event horizon \cite{Martinez:2002ru,Martinez:2004nb} and a rich family of causal structures can be obtained in this setup \cite{Martinez:2005di}. Those features are also exhibited by higher-dimensional rotating solutions even when the coupling factor is not conformal and a self-interacting potential is present \cite{Erices:2017izj}. In this line, including a self-interacting potential that induces a deviation from conformal symmetry by the presence of linear and cubic terms, allows to construct regular black holes and wormholes in (A)dS spacetime \cite{Kolyvaris:2009pc}, \cite{Anabalon:2012tu}, while the most general scalar field potential consistent with a Petrov Type-D background has been found in \cite{Anabalon:2012ta}. A charged generalization of these solutions in the static case permits the presence of a mass term for the scalar \cite{Ayon-Beato:2015ada}.

When GR is embedded in a UV complete theory, higher curvature corrections inevitably emerge. In four dimensions, quadratic higher curvature corrections are particularly interesting since they provide a power counting renormalizable theory, at the cost of introducing a ghost mode \cite{Stelle:1976gc}. Due to the topological nature of the quadratic Euler combination in four dimensions, one has only two arbitrary couplings in the quadratic sector. Unitarity is recovered if one restricts oneself to the model of the form $R^2$, since this term is dual to the Einstein-Hilbert Lagrangian with a scalar field. Consequently, it is natural to wonder how these higher curvature corrections modify the spectrum of black holes. In vacuum one can prove that, static, asymptotically flat black holes exist, only when the quadratic combination reduces to the Weyl square term of four-dimensional Conformal Gravity and the static black holes must be integrated numerically \cite{Lu:2015cqa} (see \cite{Lin:2016kip} for the asymptotically (A)dS case). Remarkably, it has been found recently that the static black hole solutions of this model can be constructed as a power series around the horizon, leading to a recurrence equation that can be solved in a closed form \cite{Podolsky:2018pfe,Svarc:2018coe,Podolsky:2019gro,Pravda:2020zno}.\\
Since the presence of a conformally coupled scalar has proven to be fruitful in the construction of analytic hairy black holes, it is also natural to wonder about how to couple in a Weyl invariant manner a scalar field with these higher curvature terms. The simplest way to perform such a task is to make use of the following tensor
\begin{equation}
        S^{\mu\nu}_{\phantom{\mu\nu}\lambda\rho} 
        = \phi^{2} R^{\mu\nu}_{\phantom{\mu\nu}\lambda\rho} -4\phi\delta^{[\mu}_{[\lambda}\nabla^{\nu]}\nabla_{\rho]}\phi
        +8\delta^{[\mu}_{[\lambda}\nabla^{\nu]}\phi\nabla_{\rho]}\phi
        -2\delta^{\mu \nu}_{\lambda \rho}\nabla_{\alpha}\phi\nabla^{\alpha} \phi \ ,\label{Stensor}
    \end{equation}
which was introduced in \cite{Oliva:2011np} in order to provide a higher dimensional generalization of the standard conformally coupled scalar field to Euler densities of higher degree (see also the works \cite{Dengiz:2011ig}-\cite{Dengiz:2012jb} for the use of a Weyl gauging in the construction of conformal higher curvature couplings with a vector field as well as a scalar). This tensor has the same symmetry properties as the Riemann tensor, and under the local Weyl rescalling 
\begin{equation}
    g_{\mu \nu}\rightarrow \Omega(x)^2g_{\mu  \nu}\ ,\ \phi\rightarrow \Omega(x)^{-1}\phi\ , \label{LWR}
\end{equation}
it transforms into
\begin{equation}
S^{\mu \nu}_{\ \ \lambda \rho}\rightarrow \Omega(x)^{-2}S^{\mu \nu}_{\ \ \lambda \rho}\ .    
\end{equation}
Conveniently, the standard Lagrangian for a conformally coupled scalar field is, up to a boundary term, equal to the second trace of the tensor \eqref{Stensor}, $S=g^{\mu \lambda}g^{\nu \rho}S_{\mu \nu \lambda \rho}$. Here after we also define its first trace as $S_{\nu\rho}=g^{\mu \lambda}S_{\mu \nu \lambda \rho}$.
\\
In this work we will consider the following model
\footnotesize
\begin{equation}
I[g,\phi]=\int\left(\frac{R-2\Lambda}{2\kappa}+\alpha_1 R^2+\alpha_2C_{\mu\nu\lambda \rho}C^{\mu\nu\lambda \rho}-\lambda\phi^4-\frac{1}{2}(\partial\phi)^2-\frac{1}{12}R\phi^2+\beta\phi^{-4}S^2\right)\sqrt{-g}d^4x \label{LAGRANGIAN}
\end{equation}
\normalsize
which can be dubbed Conformal Quadratic Gravity. We use $\kappa=8\pi G$. This theory is the natural generalization of the standard Einstein-Hilbert Lagrangian with a conformally coupled scalar field, including now conformal couplings with quadratic terms in the curvature. For $\alpha_1=\beta=0$, this model has been considered for example in \cite{Salvio:2017qkx} and \cite{Salvio:2019wcp} in the context of a cosmological model for the early universe, which may hold up to arbitrarily high energies, and has also been revisited in \cite{second} in the realm of black holes. For non-vanishing $\alpha_1$ and $\beta$,  new static, spherically symmetric, asymptotically (A)dS black holes were constructed in \cite{Caceres:2020myr}. The main purpose of the present work is to extend such explorations beyond staticity.\\
There are three algebraically independent quadratic combinations of the Riemann tensor
and the tensor $S_{\mu\nu\lambda \rho}$, namely $\left(  R_{\mu\nu\lambda \rho}R^{\mu\nu\lambda \rho}%
,R_{\mu\nu}R^{\mu\nu},R^{2}\right)  $ and $\left(  \phi^{-4}S_{\mu\nu\lambda \rho}S^{\mu\nu\lambda \rho},\phi
^{-4}S_{\mu\nu}S^{\mu\nu},\phi^{-4}S^{2}\right)  $, where each of the elements of the later set
transforms homogeneously, with a weight $\Omega^{-4}\left(  x\right)  $ under
a local Weyl rescaling (\ref{LWR}). Using the topological nature of the
Euler combination in four dimensions, it is useful to consider $R^{2}$ and
$C_{\mu\nu\lambda \rho}C^{\mu\nu\lambda \rho}$ as the basis for quadratic Lagrangians in four dimensions.
In the conformal matter sector, it can also be proven that $\phi^{-4}\left(
S_{\mu\nu\lambda \rho}S^{\mu\nu\lambda \rho}-4S_{\mu\nu}S^{\mu\nu}+S^{2}\right)  $ is a boundary term. Additionally, denoting by $W_{\mu\nu\lambda\rho}\left(  S\right)  $ the trace free part of the
tensor $S_{\mu\nu\lambda \rho}$, one can prove that $\phi^{-4}W_{\mu\nu\lambda \rho}\left(  S\right)
W^{\mu\nu\lambda \rho}\left(  S\right)  =C_{\mu\nu\lambda \rho}C^{\mu\nu\lambda \rho}$. Consequently, the Lagrangian
(\ref{LAGRANGIAN}) is the most general one containing up to quadratic terms in both
curvatures. It is interesting to notice that the tensor $S_{\mu\nu\lambda\rho}$ can be
obtained from a change of frame of the Riemann tensor, where the scalar field
emerges as a Weyl compensator. The presence of non-conformally invariant terms
in the action removes the pure gauge nature of the scalar in such construction, and allows
to introduce a new conformal degree of freedom.\\
This paper is organized as follows: In Section II we provide the field equations of the model, and review some of its properties including the existence of static black hole solutions. In Section III, we construct new solutions for this system within the C-metric ansatz as well as the Pleba\'nski-Demia\'nski family of spacetimes. Section IV is devoted to the analysis of the causal structures of some  particular cases containing Taub-NUT spacetimes as well as asymptotically AdS wormholes. In Section V, we introduce a new family of conformally invariant interactions which we dubbed Conformal K-essence, and we show that they also possess, generically, hairy black holes. Our conclusion is given in Section VI.

\section{The model and field equations}
Stationary variations of the action \eqref{LAGRANGIAN} with respect to the metric and the scalar provide the following set of field equations 
\begin{align}
    G_{\mu\nu} + \Lambda g_{\mu\nu}&=\kappa T_{\mu\nu} \\
    \left(\square-\frac{1}{6} R\right) \phi&=4 \lambda \phi^{3} -4 \beta\left(-S^{2} \phi^{-5}+\phi^{-3} R S-3 S \phi^{-4} \square \phi-3 \square\left(\phi^{-3} S\right)\right) \label{laecuacioncompleta}
\end{align}
where
\footnotesize
\begin{align}
    &T_{\mu\nu} = -2\alpha_1\left(2g_{\mu\nu} \Box R - 2 \nabla_{\nu}\nabla_{\mu}R + 2R R_{\mu\nu} - \frac{1}{2}g_{\mu\nu}R^2\right)- 4\alpha_2 \left(\nabla^{\alpha}\nabla^{\beta} C_{\alpha (\mu\nu)\beta} + R^{\alpha\beta} C_{\alpha(\mu\nu)\beta}\right) \nonumber\\
     -&2\beta\left(\left(2 R_{\mu\nu} - 2 \nabla_{\mu}\nabla_{\nu}+2g_{\mu\nu} \Box\right)(\phi^{-2}S) +12 \nabla_{(\mu}(\phi^{-3}S)\nabla_{\nu)}\phi - 6g_{\mu\nu} \nabla^{\alpha}(\phi^{-3}S\nabla_{\alpha}\phi)- \frac{1}{2}g_{\mu\nu} \phi^{-4}S^2\right)\nonumber\\
     &+\partial_{\mu} \phi \partial_\nu \phi -\frac{1}{2}g _{\mu\nu} (\partial\phi)^2 + \frac{1}{6}\left( G_{\mu\nu} -\nabla_{\mu}\nabla_{\nu}+g_{\mu\nu} \Box \right)\phi^2-\lambda g_{\mu\nu}\phi^{4} \ .
\end{align}
\normalsize
Due to the conformal invariance of the matter sector, the trace of the
energy-momentum tensor is proportional to the field equation for the scalar
and therefore, on-shell one has%
\begin{equation}
6\alpha_{1}\left(  \square-\frac{1}{6\alpha_{1}}\right)  \left(
R-4\Lambda\right)  =0\ .\label{traza}%
\end{equation}
As usual this equation signals the presence of a scalar degree of freedom
coming from the quadratic gravity terms. Such d.o.f. has an effective mass of
$m_{eff}^{2}=\left(  6\alpha_{1}\right)  ^{-1}$, and therefore, from the
perspective of flat spacetime one must
impose $\alpha_{1}>0$. Using the standard trick of multiplying and then
integrating on the spacelike section of the domain of outer communications of a
would be stationary black hole, one can prove that this equation equation
(\ref{traza}) implies the constraint $R=4\Lambda$. This argument also
applies in some supergravity models with $R^{2}$ terms \cite{Cisterna:2015iza}.\\
As a matter of fact and also for simplicity, hereafter we focus on such
sector of the space of solutions of this higher curvature theory.\\
In the absence of the term $C_{\mu\nu\lambda\sigma}C^{\mu\nu\lambda\sigma}$ in the action, i.e. when
$\alpha_{2}=0$, it was proven in \cite{Caceres:2020myr} that the model has the following
configuration as a solution%
\begin{equation}
ds^2=-f(r)dt^2+\frac{dr^2}{f(r)}+r^2d\Omega_\gamma^2 \ , \phi=\phi(r)\ ,\label{lineelement}
\end{equation}
being 
\begin{equation}
f(r)=-\frac{\Lambda}{3}r^2+\gamma\left(1-\frac{\mu}{r}\right)^2\ , \ \phi(r)=\sqrt\frac{3\mu^2(1+128\pi G\Lambda(\alpha_1+\beta))}{4G\pi}\frac{1}{r-\mu}\ ,\label{solucionneutra}
\end{equation}
and where $\gamma$ is the curvature of the horizon and the following relation between the couplings must be met
\begin{equation}
\lambda=-\frac{2\pi G\Lambda}{9(1+128\pi G\Lambda(\alpha_1+\beta))}\ .
\end{equation}
This metric, originally constructed in the model without higher derivative terms, contains a
rich set of causal structures \cite{Martinez:2002ru,Martinez:2004nb,Martinez:2005di}, which include asymptotically de Sitter black holes with a scalar field regular on and outside the horizon, as
well as asymptotically locally AdS black holes with hyperbolic horizons.
Including the higher derivative terms, and assuming staticity, this is the
only solution within the ansatz with a single blackening factor. The solution
can be charged both electrically and magnetically \cite{Caceres:2020myr}.

\section{The Pleba\'nski-Demia\'nski family}
The Pleba\'nski-Demia\'nski spacetime is the most general Petrov Type-D metric which solves the Einstein-Lambda field equations. In spite of its complexity, in \cite{Anabalon:2009qt} it was shown that such stationary spacetime does support a conformal hair on black holes as well as gravitational stealths (non-trivial scalar field profiles with vanishing energy-momentum tensor \cite{AyonBeato:2004ig,AyonBeato:2005tu}) in the prototypical second order scalar-tensor model $\alpha_1=\alpha_2=\beta=0$, while the the static C-metric ansatz was also shown to lead to non-trivial solutions in this system in \cite{Charmousis:2009cm}.\\
The ansatz leading to the Pleba\'nski-Demia\'nski solution in General Relativity
\cite{Debever} can be written as \cite{Griffiths:2005qp}
\footnotesize
\begin{equation}
ds^{2}=\frac{1}{\left(  1-\alpha xy\right)  ^{2}}\left[  -\frac{X\left(
x\right)  \left(  d\tau-\omega y^{2}d\sigma\right)  ^{2}}{x^{2}+\omega
^{2}y^{2}}+\frac{Y\left(  y\right)  \left(  \omega d\tau+x^{2}d\sigma\right)
^{2}}{x^{2}+\omega^{2}y^{2}}+\left(  x^{2}+\omega^{2}y^{2}\right)  \left(
\frac{dy^{2}}{Y\left(  y\right)  }+\frac{dx^{2}}{X\left(  x\right)  }\right)
\right]  \ .\label{plebdem}%
\end{equation}
\normalsize
Here, $\alpha$ and $\omega$ are parameters introduced in
\cite{Griffiths:2005qp}, which facilitate taking different limits for physically
relevant cases. Since our matter field is conformally invariant, the trace of the field
equations leads to equation (\ref{traza}), and inspired by the no-hair result described above, we solve such an equation by imposing
\begin{equation}
R=4\Lambda\ .
\end{equation}
This equation is in particular solved by quartic polynomials on their variables for $X$
and $Y$, while in this sector, the remaining field equations imply the following non-trivial
solution of the system%
\begin{align}
X\left(  x\right)    & =-\left(  \alpha^{2}y_{0}+\frac{1}{3}\Lambda
_{0}\right)  x^{4}-y_{2}x^{2}+x_{0}\label{polX}\ , \\
Y\left(  y\right)    & =-\left(  x_{0}\alpha^{2}+\frac{1}{3}\Lambda_{0}%
\omega^{2}\right)  y^{4}+y_{2}y^{2}+y_{0}\label{polY}\ ,\\
\phi\left(  x,y\right)    & =\frac{B\left(  1-\alpha xy\right)  }{\left(
1+\alpha xy\right)  }\label{phi}\ ,%
\end{align}
which solves the field equations provided%
\begin{equation}
\kappa B^{2}=6\left(1+16\kappa\Lambda_{0}\left(  \alpha_{1}+\beta\right)  \right)\ \ 
\text{and}\ \  \lambda=-\frac{\Lambda_0}{6B^2}.
\end{equation}
This configuration extends to the realm of Conformal Quadratic Gravity the hairy Pleba\'nski-Demia\'nski solution constructed in \cite{Anabalon:2009qt} as well as the C-metric found in \cite{Charmousis:2009cm}. As in \cite{Anabalon:2009qt}, there are also extra branches that lead to non-trivial
scalar field profiles with vanishing energy-momentum tensor on locally
constant curvature spacetimes. Nevertheless,
the configuration defined by (\ref{polX})-(\ref{phi}) defines a spacetime with non-trivial curvature. Here, $x_0, y_0$ and $y_2$ are integration constants. Along the lines of \cite{Griffiths:2005qp} one can analyze the different causal structures contained in \eqref{plebdem}. Below, we obtain the C-metric as well as the topological Taub-NUT spacetime integrating the equations from scratch, also we provide comments on their causal structures.  

\subsection{C-metric black holes}

The metric%
\begin{equation}
ds^{2}=\frac{1}{\left(  y-Ax\right)  ^{2}}\left[  \frac{dx^{2}}{X\left(
x\right)  }+\frac{dy^{2}}{Y\left(  y\right)  }-Y\left(  y\right)
dt^{2}+X\left(  x\right)  d\sigma^{2}\right]  \ ,
\end{equation}
with%
\begin{align}
X(x)  & =-A^{2}y_{4}x^{4}-Ay_{3}x^{3}-y_{2}x^{2}-\frac{y_{3}(4y_{2}y_{4}%
-y_{3}^{2})}{8Ay_{4}^{2}}x+x_{0}\ ,\\
Y(x)  & =y_{4}y^{4}+y_{3}y^{3}+y_{2}y^{2}+\frac{y_{3}(4y_{2}y_{4}-y_{3}^{2}%
)}{8y_{4}^{2}}y-A^{2}x_{0}-\frac{\Lambda_{0}}{3}\ ,\\
\phi(x,y)  & =4y_{4}\sqrt{\frac{3}{2\kappa}\left(1+16\kappa\Lambda_{0}\left(  \alpha_{1}%
+\beta\right)  \right)  }\frac{y-Ax}{2y_{4}\left(  Ax+y\right)  +y_{3}}\ ,
\end{align}
\bigskip is a solution of the higher-derivative action \eqref{LAGRANGIAN}, provided the
couplings of the theory are related as%
\begin{equation}
\lambda=-\frac{\Lambda_{0}\kappa}{36\left(  1+16\kappa\Lambda_{0}\left(  \alpha
_{1}+\beta\right)  \right)  }\ \label{relation}.
\end{equation}
Here $x_{0},y_{2},y_{3}$ and $y_{4}$ are constants.
Depending on the structure of zeros of the polynomial $X\left(  x\right)  $
one can have constant $t$ surfaces with different topologies. Our solution is
an extension to the higher derivative regime of the hairy C-metric found in
\cite{Charmousis:2009cm} and \cite{Anabalon:2009qt}. The sign of the integration constants is not fixed a priori,
in contraposition to what occurs with a term of the form $Q^{2}$ in the metric,
when $Q$ is the electric charge. When $X\left(  x\right)  $ has two zeros at $x=x_{\max}$ and $x=x_{\min}$, and
it is positive between such points, one can restrict the coordinate $x$ to such regions and
identify the coordinate $\sigma$. In general, this introduces conical
singularities at $x=x_{\max}$ and $x=x_{\min}$, both of which can be remarkably removed in
this case due to the backreaction of the conformal scalar, leading to the
cohomogeneity-two black holes reported in \cite{Anabalon:2009qt}. As shown above,
such structure persists when higher curvature terms are included.

\section{Topological Taub-NUT with a conformal hair}

It is known from \cite{Anabalon:2009qt} that a limit can be taken from the
Pleba\'nski-Demia\'nski spacetime which leads to Taub-NUT spacetime in the second
order theory defined by (\ref{LAGRANGIAN}) with $\alpha_{1}=\alpha_2=\beta=0$. Even more, such
spacetimes can be conveniently obtained by direct integration \cite{Bhattacharya:2013hvm} and by using a generalization of the Ernst
generating technique \cite{Bardoux:2013swa,Astorino:2014mda}. In presence of the higher
derivative terms, when the relation \eqref{relation} holds, the configurations are corrected and lead to%
\footnotesize
\begin{equation}
ds^{2}=-f\left(  r\right)  \left(  dt+n\left\{
\begin{array}
[c]{c}%
4\sin^{2}\left(  \frac{\theta}{2}\right)  \\
\rho^{2}\\
4\sinh^{2}\left(  \frac{\theta}{2}\right)
\end{array}
\right\}  d\phi\right)  ^{2}+\frac{dr^{2}}{f\left(  r\right)  }+\left(
r^{2}+n^{2}\right)  \left(  \left\{
\begin{array}
[c]{c}%
d\theta^{2}\\
d\rho^{2}\\
d\psi^{2}%
\end{array}
\right\}  +\left\{
\begin{array}
[c]{c}%
\sin^{2}\theta\\
\rho^{2}\\
\sinh^{2}\psi
\end{array}
\right\}  d\phi^{2}\right)  \ ,\label{metriccolumna}%
\end{equation}
\normalsize
with%
\begin{align}
f\left(  r\right)    & =-\frac{\Lambda_{0}}{3}\left(  r^{2}+n^{2}\right)
+\left(  \gamma-\frac{4}{3}n^{2}\Lambda_{0}\right)  \frac{\left(  r-M\right)
^{2}}{r^{2}+n^{2}}\label{fnut}\\
\phi\left(  r\right)    & =2\sqrt{\frac{3\left(  1+16\kappa\Lambda_{0}\left(
\alpha_{1}+\beta\right)  \right)  }{2\kappa}}\frac{\sqrt{M^{2}+n^{2}}}%
{r-M}\label{escalarnut}%
\end{align}
where $\gamma=+1,0,-1$, corresponds to the first, second and third line of
(\ref{metriccolumna}), respectively, $M$ is the mass and $n$ the NUT parameter (notice that Taub-NUT/Bolt solutions with scalar hair have also been recently obtained in Ref. \cite{Arratia:2020hoy}).\\
Restricting to the hyperbolic case ($\gamma=-1$), one can perform the following formal
transformation on such a configuration%
\begin{equation}
t=ui+nT\text{, }\psi=\theta+\frac{\pi}{2}i\text{ and }\phi=T\ ,\label{transf}
\end{equation}
which leads to%
\begin{equation}
ds^{2}=\frac{dr^{2}}{f\left(  r\right)  }+f\left(  r\right)  \left(
du+2n\sinh\theta dT\right)  ^{2}+\left(  r^{2}+n^{2}\right)  \left(
-\cosh^{2}\theta dT^{2}+d\theta^{2}\right) \label{wormhole} \ .
\end{equation}
with $f(r)$ and the scalar field given by \eqref{fnut} and (\ref{escalarnut}), respectively. For negative
cosmological constant, when $f\left(  r\right)  >0$ in the domain
$-\infty<r<\infty$, a condition that is achieved in an open set of the parameter
space $(n,M)$, this configuration describes the hairy extension of the
asymptotically locally AdS wormhole found in \cite{Anabalon:2018rzq}. When a
Maxwell field is included, this spacetime can be embedded in $\mathcal{N}=2$
gauged supergravity and admits non-trivial Killing spinors \cite{Anabalon:2020loe}. It is interesting to notice, that the formal transformation \eqref{transf} is not a double Wick rotation but still it leads to a Lorentzian geometry with a new causal structure. It is known that the hyperbolic Taub-NUT AdS is devoid of closed timelike curves, and the wormhole geometry and its hairy extension presented here maintain this property, which was proven in \cite{Anabalon:2018rzq} along the lines of the analysis performed in \cite{Coussaert:1994tu} for a particular quotient of AdS$_3$ spacetime.

\section{A general family of conformally invariant Lagrangians: Conformal
K-essence}

Since under a local Weyl rescalling the tensor $S_{\ \ \lambda\rho}^{\mu\nu}$ transforms as $S_{\ \ \lambda\rho}^{\mu\nu}\rightarrow\Omega
^{-4}S_{\ \ \lambda\rho}^{\mu\nu}$, the matter theory%
\begin{equation}
I_{\text{matter}}=\int d^{4}x\sqrt{-g}\phi^{4}F\left(  \phi^{-4}S\right)
\ ,\label{genf}%
\end{equation}
is invariant under conformal transformations for an arbitrary function
$F\left(  Y\right)  =F\left(  \phi^{-4}S\right)  $. When $F$ is a constant
function, the action (\ref{genf}) corresponds to the conformal potential,
while for $F$ being the identity function, (\ref{genf}) reduces to the
standard conformally coupled scalar field, up to a boundary term. The
combination $Y=\phi^{-4}S=\phi^{-4}\left(  \phi^{2}R-6\phi\square\phi\right)
$ is indeed conformally invariant. The action defined by (\ref{genf}) can
be thought of as a conformally invariant extension of a K-essence Lagrangian,
namely a conformally invariant version of the action $K\left(  \phi\square
\phi\right)  $. Coupling this conformal invariant action to GR%
\begin{equation}
I=\int\sqrt{-g}d^{4}x\left[  R-2\Lambda+\phi^{4}F\left(  \phi^{-4}S\right)
\right]  \ ,\label{fullF}%
\end{equation}
leads to the field equations%
\begin{align}
G_{\mu\nu}+\Lambda g_{\mu\nu} &  =T_{\mu\nu}\ \\
2\phi^{3}F\left(  Y\right)  -R\phi\frac{dF}{dY}+9\square\phi\frac{dF}%
{dY}-3\square\left(  \phi\frac{dF}{dY}\right)   &  =0\ \label{fieldF}%
\end{align}
where%
\footnotesize
\begin{align}
T_{\mu\nu} &  =\frac{1}{2}g_{\mu\nu}\phi^{4}F\left(  Y\right)  -\frac{dF}{dY}\left(
\phi^{2}R_{\mu\nu}-6\phi\nabla_{\mu}\nabla_{\nu}\phi\right)  +\nabla_{\mu}\nabla
_{\nu}\left(  \frac{dF}{dY}\phi^{2}\right)  \nonumber \\
&-3\left(  \nabla_{\mu}\left(
\frac{dF}{dY}\phi\nabla_{\nu}\phi\right)  +\nabla_{\nu}\left(  \frac{dF}{dY}%
\phi\nabla_{\mu}\phi\right)  \right) -\square\left(  \frac{dF}{dY}\phi^{2}\right)  g_{\mu\nu}+3\nabla^{\lambda}\left(
\frac{dF}{dY}\phi\nabla_{\lambda}\phi\right)  g_{\mu\nu}\ .\label{Tmunu}%
\end{align}
\normalsize
One can check that, as expected from conformal invariance, the trace of the energy-momentum tensor (\ref{Tmunu}) vanishes by virtue of the equation
for the scalar field (\ref{fieldF}).
The configuration%
\begin{align}
ds^{2}  & =-\left(  -\frac{\Lambda_{0}}{3}r^{2}+\left(  1-\frac{M}{r}\right)
^{2}\right)  dt^{2}+\frac{dr^{2}}{-\frac{\Lambda_{0}}{3}r^{2}+\left(
1-\frac{M}{r}\right)  ^{2}}+r^{2}\left(  d\theta^{2}+\sin^{2}\theta d\psi
^{2}\right)  \label{MTZmetric}\\
\phi\left(  r\right)    & =\frac{A}{r-M}\ ,\label{MTZField}%
\end{align}
leads to a constant Weyl invariant combination
\begin{equation}
Y=\phi^{-4}S=\phi^{-4}\left(  \phi^{2}R-6\phi\square\phi\right)
=\frac{4\Lambda_{0}M^{2}}{A^{2}}\ .
\end{equation}
Here we focus on the $\gamma=1$ case. Specifying the function $F$ in the action (\ref{fullF}), the field equations
on the configuration defined by (\ref{MTZmetric}) and (\ref{MTZField}) lead to
the following two constraints%
\begin{align}
2M^{4}\Lambda_{0}+A^{4}F\left(  Y\right)    & =0\ ,\ \label{const1}\\
A^{2}F\left(  Y\right)  -2\Lambda_{0}M^{2}\frac{dF\left(  Y\right)  }{dY}  &
=0\ .\label{const2}%
\end{align}
These constraints can be interpreted as follows: the constraint (\ref{const1})
fixes the value of the constant appearing on the scalar field $A$, while the
constraint (\ref{const1}) fixes the different parameters that appear in the
action functional through the election of the function $F\left(  Y\right)  $.
For example, choosing $F\left(  Y\right)  =-\frac{1}{12}Y-\lambda$ in
(\ref{fullF}) leads to the standard conformally coupled scalar field with a
conformal potential coupling $\lambda$ such that
\begin{equation}
\phi\left(  r\right)  =2\sqrt{3}\frac{M}{r-M}\text{ and }\lambda
=-\frac{\Lambda_{0}}{72}\ ,
\end{equation}
which reduces to the solution found in \cite{Martinez:2002ru} with
the normalization $16\pi G=1$.
We have therefore proved that provided the constraints (\ref{const1}) and
(\ref{const2}) are fulfilled, the black hole configuration  (\ref{MTZmetric})
and (\ref{MTZField}) is a solution of General Relativity supported by a
general Conformal K-essence field (\ref{genf}).

\section{Further comments}
We have shown that the Pleba\'nski-Demia\'nski spacetime, and therefore all Type-D metrics, as for example the C-metric and topological Taub-NUT spacetimes can be embedded as solutions of a quadratic gravity model, conformally coupled to a scalar field with higher derivative corrections. Some of these spacetimes have already been studied in the presence of a conformally coupled scalar field with second order dynamics and we expect that in the black hole case, the higher derivative terms may induce modifications on the thermodynamics, as well as in the perturbative stability properties of the solutions. Our solutions can also be extended to include a Maxwell field including both electric and magnetic sources, which in the case of static black holes with hyperbolic horizons in AdS may describe the causal structure of a black hole inside a black hole \cite{Martinez:2005di}.\\
The tensor $S_{\ \ \lambda\rho}^{\mu\nu}$ is the building block that allowed to construct conformal couplings of a scalar to Euler densities in arbitrary dimension \cite{Oliva:2011np}. In such setup, one can also construct black hole solutions with simple scalar field profiles, provided the couplings of the different terms in the matter sector fulfil a relation \cite{Giribet:2014bva,Chernicoff:2016jsu,Chernicoff:2016uvq,Chernicoff:2016qrc}, leading to an interesting phase structure \cite{Giribet:2014fla,Galante:2015voa,Hennigar:2015wxa,Hennigar:2016xwd}. Such constraint is of a different sort than the one imposed by \eqref{relation} which relates the gravitational and the matter couplings. Using the same tensor, it would be interesting to construct new higher curvature and derivative couplings in higher dimensions and classify them. One may even consider complete contractions of tensor products of Weyl tensor $C_{\ \ \lambda\rho}^{\mu\nu}$ and copies of the tensor $S_{\ \ \lambda\rho}^{\mu\nu}$. Such combinations will transform homogeneously under conformal transformations and therefore it will lead to conformally invariant Lagrangians by the introduction of a compensating power of the scalar field. In order to go beyond the standard conformally coupled scalar field, but still within the realm of a second order theory, one is forced to introduce a second scalar leading to a Bi-scalar Extensions of Horndeski Theories \cite{Charmousis:2014zaa}, which may admit some of the stationary solutions we have found in this paper.\\
Finally, we have introduced a general conformally invariant model that depends on an arbitrary function, which we have dubbed Conformal K-essence. It is tempting to conjecture that such general family of theories do also admit stationary solutions in the Pleba\'nski-Demia\'nski family. To study different application of such Conformal K-essence it would be interesting to study first the potential formation of caustic as was done for example in \cite{Babichev:2016hys} for the standard K-essence theory. Work along these lines is in progress.

\section{Acknowledgments} 
We thank Andrés Anabalón, Nicolás Cáceres, José Figueroa, Nicolás Mora, Marcelo Oyarzo and Ricardo Stuardo for valuable discussions.  We would like to express our gratitude to Carmen Lagler for her detailed proofreading of this manuscript. This work is partially funded by Beca Chile de Postdoctorado 74200012 and FONDECYT grant 1181047 and 1210500. J.O. also thanks the support of Proyecto de Cooperaci\'on Internacional 2019/13231-7 FAPESP/ANID.

\end{document}